\documentclass[aps,prl,twocolumn]{revtex4}

\usepackage{amsmath}
\usepackage{amsfonts}
\usepackage{amssymb}
\usepackage{epsfig}
\usepackage{epstopdf}
\usepackage[usenames]{color}

\newcommand{\ket}[1]{\mbox{$|#1\rangle$}}

\begin{document}
\title{Violation of the Leggett-Garg inequality with weak measurements of photons}
\author{M. E. Goggin$^{1,2}$, M. P. Almeida$^1$, M. Barbieri$^{1,3}$, B. P. Lanyon$^1$,  J. L. O'Brien$^4$,  A. G. White$^1$, G. J. Pryde$^5$}
\affiliation{$^1$ Department of Physics and Centre for Quantum Computer Technology, University of Queensland, Brisbane 4072, Australia\\
$^2$ Physics Department, Truman State University, Kirksville, MO, 63501, USA\\
$^3$  Groupe d'Optique Quantique, Laboratoire Charles Fabry, Institut d'Optique - Graduate School, F91127 Palaiseau Cedex, France\\
$^4$ Centre for Quantum Photonics, H. H. Wills Physics Laboratory \& Department of Electrical and Electronic Engineering, University of Bristol, Merchant Venturers Building, Woodland Road, Bristol, BS8 1UB, UK\\
$^5$  Centre for Quantum Dynamics and Centre for Quantum Computer Technology, Griffith University, Brisbane, 4111, Australia}

\begin{abstract}
By weakly measuring the polarization of a photon between two strong polarization measurements, we experimentally investigate the correlation between the appearance of anomalous values in quantum weak measurements, and the violation of realism and non-intrusiveness of measurements. A quantitative formulation of the latter concept is expressed in terms of a Leggett-Garg inequality for the outcomes of subsequent measurements of an individual quantum system. We experimentally violate the Leggett-Garg inequality for several measurement strengths.  Furthermore, we experimentally demonstrate that there is a one-to-one correlation between achieving strange weak values and violating the Leggett-Garg inequality.

\end{abstract}
\maketitle

One of the fundamental challenges in understanding the world is addressing the question of whether or not systems have a real state independent of observation. Key ideas addressing this question include the Bell inequality, with its joint assumption of realism and/or locality \cite{Bell1, Bell2,Vienna2007}, and contextuality tests, which examine whether or not identical experiments produce results in different ``classically equivalent" contexts \cite{contextuality}. Both examples rely on state preparations---which can be viewed as the outcome of an initial measurement---followed by a final measurement from which outcomes are computed. A conceptually elegant extension to this is the idea of measuring the system between the initial and final measurements. A test---the Leggett-Garg inequality (LGI) --- has been proposed which is based on a series of three measurements, and its violation implies a failure of the assumptions of realism and noninvasive detection \cite{Leggett1985}. Here we present a photonic experiment that violates a generalised LGI  \cite{Jordan2006} by 14 standard deviations.  We used a weak measurement to minimize invasiveness \cite{weakmeas, anomalous}. Furthermore, we experimentally demonstrate a one-to-one relation \cite{Williams2008} between LGI violations and strange weak-valued measurements, which are also connected with the inability to determine a classical trajectory between an earlier and a later measurement.

Quantum mechanics only provides outcome probabilities, without describing the state in between the preparation and the measurement.  In the standard view, a measurement interrupts the unitary evolution of the state via the collapse of its wavefunction.  In 1985, Leggett and Garg introduced an inequality \cite{Leggett1985} bounding the correlation for consecutive measurements on a single system based on the joint assumptions of macroscopic realism and noninvasive measurement: a macroscopic system will at all times be in only one of its available states, and it is possible to determine such a state with arbitrarily small disturbance on its subsequent evolution. Thus a violation of the LGI tells us that either earlier measurements perturbed the system, changing the results of later measurements, or macroscopic realism is untenable, or both.

The trick to testing the LGI is to be able to monitor the system without ``collapsing'' the wavefunction: to make a so-called weak measurement \cite{weakmeas}. A weak measurement is one for which it is possible, in principle, to reduce the back action on the system to an arbitrarily small amount.

Shortly after Leggett and Garg introduced their inequality, Aharanov and Vaidman suggested that observing the result of a weak measurement conditioned on a specific result of a separate projective measurement leads to unusual results. One unusual property, dubbed  \emph{strange} weak values, is that the value assigned in this way may lie outside the eigenspectrum of the operator being measured \cite{anomalous}. Although such strange weak values have been observed \cite{steinberg,opt1,3box,ours, onur}, the idea that the measured value lies outside the operator's eigenspectrum raises questions about realistic descriptions of the system's state, and of the measurement process \cite{leggett1989, peres}. In this sense, strange weak values explore the same concepts (or raise the same problems) as the LGI. A formal connection between a generalized LGI and weak values has been recently proposed \cite{Williams2008}, Specifically that there is a one-to-one relationship between the violation of the LGI and the measurement of a strange weak value. That is, the LGI is violated if and only if the experiment yields a strange weak value.

Although formulated in the context of a macroscopic system, when applied to a single photon the LGI tests the presence of underlying definite trajectories for the evolution of a quantum system. The question is actually whether our experiment, in which a single photon is prepared, weakly measured and eventually post-selected, could be understood using the notions of objective existence of measured properties and noninvasiveness of the measurement. Here we show an example in which at least one of these assumptions fails: we need to admit that either the photon has no deterministic evolution from the input to the output state, since it is not in a well-defined state, or that such an evolution cannot be monitored, however weak the monitor, without being irremediably compromised.

The generalised Leggett-Garg correlation function\cite{Jordan2006} is
\begin{equation}
B=\langle\mathcal{M}_a\mathcal{M}_b\rangle+\langle\mathcal{M}_b\mathcal{M}_c\rangle-\langle\mathcal{M}_a\mathcal{M}_c\rangle,
\label{eq:LGP}
\end{equation}
where $\{\mathcal{M}_a, \mathcal{M}_b, \mathcal{M}_c\}$ is a set of three consecutive weak measurements on the photon polarization.  For a physical system with the properties of both realism and noninvasive detection, $B$ must satisfy $-3 \leqslant B \leqslant 1$ which is the Leggett-Garg Inequality (LGI). On the other hand, quantum measurements produce values of $B$ satisfying \cite{Jordan2006}: $-3 \leqslant B \leqslant 1.5$.  Thus, there is a violation of the LGI if $1 < B \leqslant 1.5$.  Detailed analysis\cite{Williams2008} shows that, according to quantum mechanics, the violation of the LGI does not depend on the measurement strength of either $\mathcal{M}_a$ or $\mathcal{M}_c$, so we can perform strong projective measurements in those cases without loss of generality.  

\begin{figure}[th]
\includegraphics[viewport= 70 125 750 590, clip, width= \columnwidth]{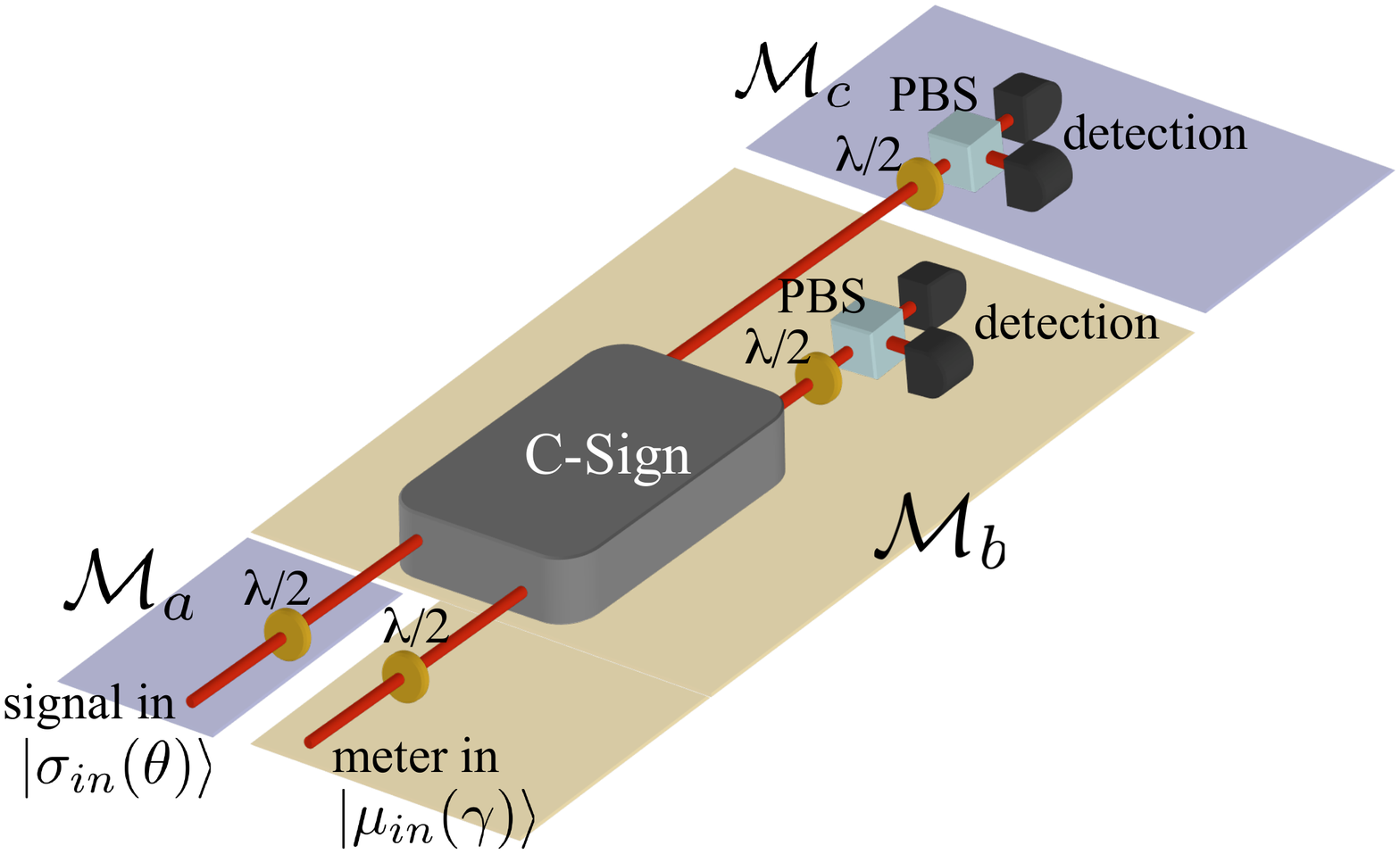}
\caption{Conceptual representation of the experiment. A single photon is input into the left (signal) arm of the apparatus, where it is prepared in a state using a half waveplate ($\lambda /2$). A weak measurement of the polarization is made by interacting the photon with a second (meter) photon in a weak measurement device, which operates via measurement-induced nonlinearity \cite{Pryde2004,Pryde2005a,OBrien2003,OBrien2003,CnotDe,CnotJp}. The signal and meter photon are then measured in coincidence. A coincidence count flags successful operation. The signal and meter photons are produced in pairs by spontaneous parametric downconversion from a bismuth triborate (BiBO) crystal, pumped by 100fs pulses from a frequency doubled Ti:Sa laser centred around 820nm. See the main text for a description of $\mathcal{M}_a$, $\mathcal{M}_b$, and $\mathcal{M}_c$}
\label{fig:setup}
\end{figure}

A conceptual schematic of our experimental setup is shown in Fig. \ref{fig:setup}. We input a photon in the state $\ket{\sigma_{in}}=\cos\tfrac{\theta}{2} \ket{H}+\sin\tfrac{\theta}{2}\ket{V}$ where $\ket{H}$ and $\ket{V}$ represent horizontal and vertical polarisation states. We consider $\ket{\sigma_{in}}$ as the state prepared by the first measurement $\mathcal{M}_a$; in this way, we can deterministically assign the value 1 to $\mathcal{M}_a$.  The weak measurement, $\mathcal{M}_b$, measures the $\mathcal{S}_1$ Stokes parameter of a single photon \cite{Pryde2005a}, corresponding to the degree of polarization in the horizontal-vertical basis. The final measurement, $\mathcal{M}_c$, is the $\mathcal{S}_2$ Stokes parameter which quantifies the degree of polarisation in the diagonal-antidiagonal basis. Thus,  in our case we can rewrite \eqref{eq:LGP} as
\begin{equation}
B=\langle\mathcal{S}_1\rangle+\langle\mathcal{S}_1\mathcal{S}_2\rangle-\langle\mathcal{S}_2\rangle,
\label{eq:LG}
\end{equation}
Note that $\mathcal{S}_1$ is a weak measurement in the sense of Leggett and Garg's original paper \cite{Leggett1985} and $\mathcal{S}_2$ is a projective measurement. 


The weak measurement is implemented using a nondeterministic linear optical controlled-sign gate \cite{Pryde2004,OBrien2003,OBrien2004,CnotDe,CnotJp}. The signal qubit $\ket{\sigma_{\textrm{in}}}$ in the gate's control mode becomes entangled with a meter qubit in the target mode, and this latter qubit is measured to obtain the weak measurement result. The strength of the measurement can be set by changing the meter input state $\ket{\mu_{\mathrm{in}}}=\gamma \ket{D} + \bar{\gamma}\ket{A}$ (with $\ket {D}=\tfrac{1}{\sqrt{2}}\left(\ket H + \ket V \right)$, $\ket {A}=\tfrac{1}{\sqrt{2}}\left(\ket H - \ket V \right)$, $\gamma^2+\bar{\gamma}^2=1$, and $\gamma \in \mathbb{R}$ without loss of generality). The strength of the measurement is quantified by the \textit{knowledge} \cite{Pryde2004}, $K=2\gamma^2-1$. When $K=1$ ($\gamma=1$), full information is extracted about the
measured photon, while when $K=0$ ($\gamma=1/\sqrt{2}$), the photon is left completely undisturbed and unmeasured. Between those two extremes, the photon undergoes partial disturbance, and one can extract the average value of $\mathcal{S}_1$ as \cite{Pryde2005a}:
\begin{equation}
{\langle\hat{\mathcal{S}}_1\rangle}=\frac{P(D)-P(A)}{K},
\label{eq:weakmeas}
\end{equation}
where $P(D)$ and $P(A)$ are the probabilities of measuring the state $D$ and $A$ of the meter photon, respectively. The weak character is reflected in the fact that an increasing number of copies are necessary to estimate $\langle\hat{\mathcal{S}_1\rangle}$ to a given precision as $K$ decreases.

The \emph{weak value} \cite{weakmeas, anomalous} is calculated by post-selecting on a measurement of $\mathcal{M}_c$ that results in the diagonal state for the signal photon \cite{Pryde2004}. In our implementation, we considered the weak value:
\begin{equation}
{_D\langle\hat{\mathcal{S}}_1\rangle}=\frac{P(D|D)-P(A|D)}{K},
\label{eq:weakval}
\end{equation}
where $P(A|D)$ is the probability of measuring the state $A$ of the meter photon, conditioned on the post-selection of the signal photon in state $D$, and similarly for $P(D|D)$.  We expect anomalous values to emerge when $\ket{\sigma_{\textrm{in}}}$ approaches the antidiagonal state  $\ket {A}$.


\begin{figure}[th]
\includegraphics[viewport=50 50 600 750, width= \columnwidth]{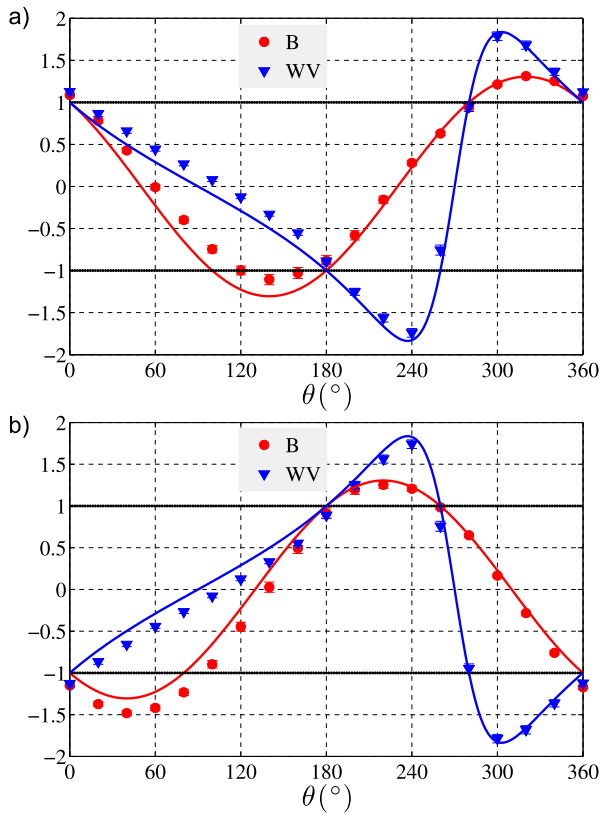}
\caption{Variation of the weak value (WV ($\blacktriangledown$) ) and the Leggett-Garg parameter (B ($\bullet$)) for a range of input states $|\sigma_{\textrm{in}}\rangle$ parametrised by $\theta$. Post-selection is on the diagonal state $|D\rangle$, and the measurement strength is $K=0.5445 \pm 0.0083$. Error bars arise from poissonian counting statistics. (a) B parameter formed with $\mathcal{M}_b=\mathcal{S}_1$. (b) B parameter formed with $\mathcal{M}_b=-\mathcal{S}_1$. Solid lines are the theoretical predictions from Eqs. \ref{eq:weakval} and \ref{eq:LG}. The horizontal lines at $\pm 1$ indicate the limits of the eigenspectrum of $\mathcal{S}_1$. $B_{\mathrm{max}}=1.312 \pm 0.022$}
\label{fig:data1}
\end{figure}

In Fig. 2a, we report the measured values for the LG correlation function, $B$, and the weak values for the choice ${K=0.5445\pm0.0083}$ as $\theta$ is varied in the range $[0,2\pi]$~\cite{WVfootnote}. We attribute the discrepancies between the theoretical curves and the data points to imperfect quantum interference at the gate, introducing mixture in the states. At this measurement strength we achieve a maximum value of the LG correlation function of $B_{\mathrm{max}}=1.312 \pm 0.022$.  It is clear that a violation of the LGI and the appearance of anomalous positive weak values are manifestly correlated; on the other hand, we also have strange negative weak values corresponding to a value of $B$ compatible with Leggett and Garg's hypotheses. This seemingly contradicts the results in \cite{Williams2008}. 
We can restore the one-to-one correspondence if we consider the case $\mathcal{M}_b=-\mathcal{S}_1$, shown in Fig. 2b.  A violation of the LGI is observed in correspondence with the anomalous values, in the same range where we had none previously. This happens because quantum mechanics predicts no values for $B$ below the classical limit of $-3$: a violation is possible only by exceeding the upper bound, hence the need for anomalous positive values.


\begin{figure}[th]
\includegraphics[viewport=50 50 600 750, width= \columnwidth]{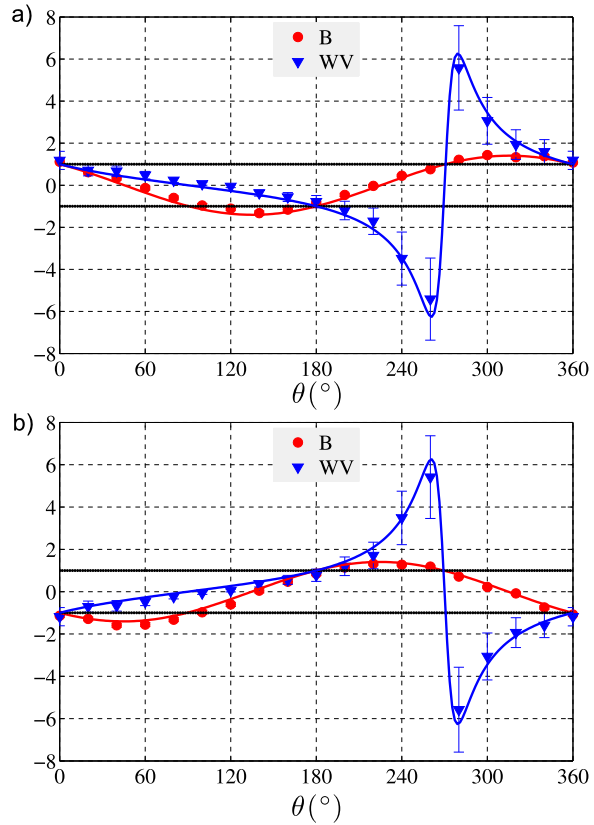}
\caption{Same as Fig. \ref{fig:data1} except the measurement strength is $K=0.1598 \pm 0.0091$. Error bars are larger than in Fig. \ref{fig:data1} because of the reduced post-selection probability. $B_{\mathrm{max}}=1.436 \pm 0.053$}
\label{fig:data2}
\end{figure}

To understand the effect of the measurement strength on violation of the LGI, we repeated the experiement with $K=0.1598\pm0.0091$. Here we achieve a maximum value of the LG correlation function of $B_{\mathrm{max}}=1.436 \pm 0.053$.  We also observe a one-to-one correlation between LGI violation and strange weak values for this measurement strength as shown in Fig. \ref{fig:data2}. Note that we also observe weak values further outside the spectrum of $\mathcal{S}_1$ but with larger error bars as a result of a reduced post-selection probability. For this condition, we would expect the visibility in the $D-A$ basis to be modified from its maximum value by a factor of $\sqrt{1-K^2}\approx0.98$. For comparison, this factor is equal to 0.84 when $K=0.5445$ as in Fig. 2.  Thus, the measurement disturbs the photon significantly less when $K=0.1598$ than when $K=0.5445$. 

\begin{figure}[th]
\includegraphics[viewport=50 50 850 550, width= \columnwidth]{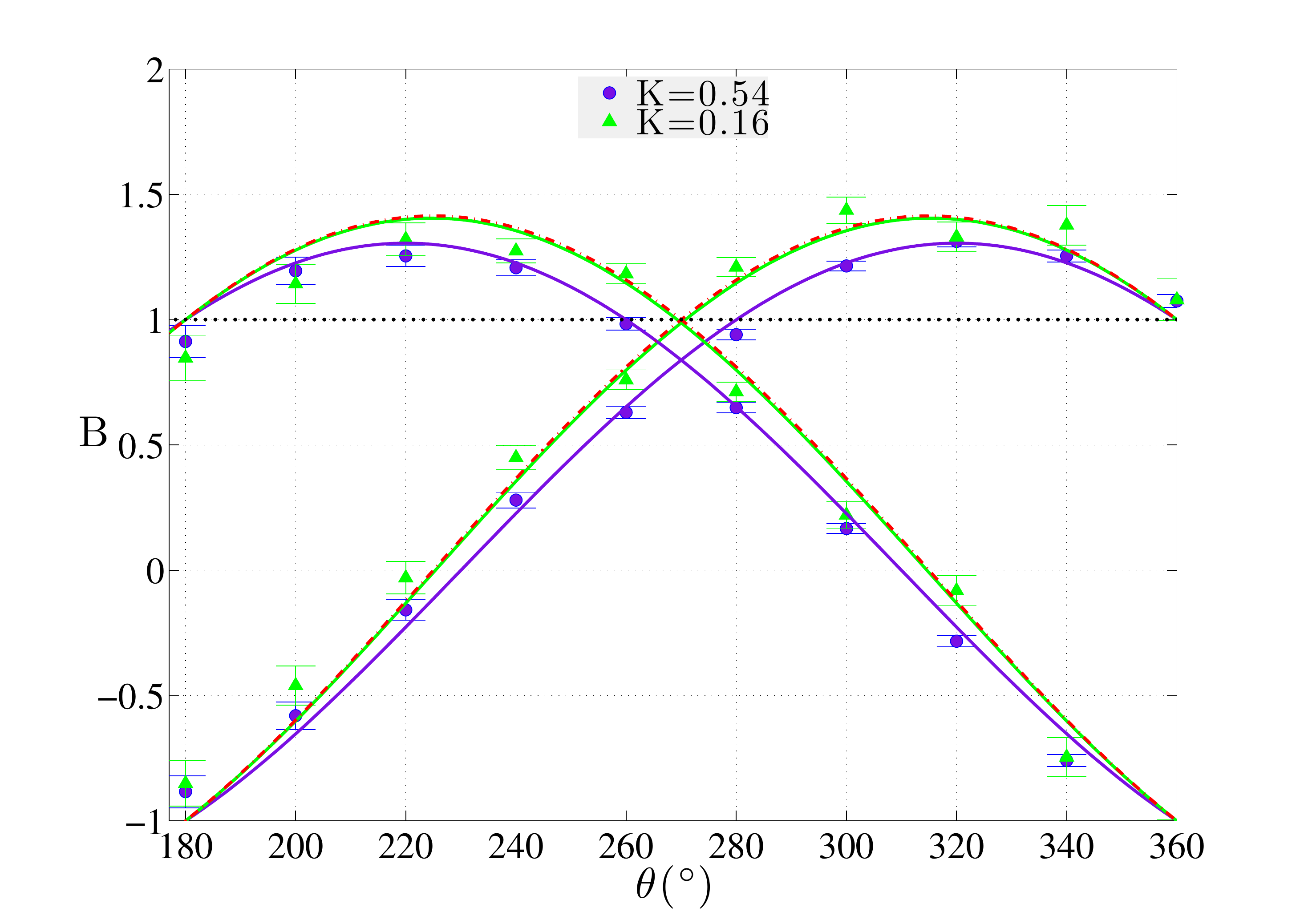}
\caption{Comparison of the Leggett-Garg correlation function (B) for a range of input states $|\sigma_{\textrm{in}}\rangle$, for the two measurement strengths studied above, $K=0.54$ ($\bullet$) and $K=0.16$ ($\blacktriangle$). It is shown that the range of violation of the LGI is larger when the measurement is less intrusive.  The red dashed line is the theoretical result for a zero strength measurement.}
\label{fig:data3}
\end{figure}

Although violation of the LGI and the observation of strange weak values give essentially the same information, the violation of the LGI has to be regarded in some way as a more robust indicator. The LGI involves all the measured photons, while the measurement of anomalous weak values only uses a subset of the measured photons -- those that satisfy the post-selection criterion.  Therefore, the probability of detecting a photon with an anomalously large weak value becomes fainter as we decrease the measurement strength.  By adjusting the measurement strength, we were able to verify that the peak violation of the LGI increases as the measurement strength decreases \cite{Jordan2006}.  We also observed that the range of $\theta$ values for which a violation occurs is extended when the measurement is weaker (see Fig. \ref{fig:data3}). Physically, the dependence of the LGI violation on measurement strength can be understood in the following way. As the measurement gets stronger the quantum system becomes more ``realistic'' in the sense that the photon is ``prepared'' in a state by the measurement and will have that value until measured again.  Therefore the system will satisfy the LGI.  But as the measurement gets weaker, the quantum evolution is less disturbed and \emph{apparently} less realistic thus violating the LGI. 


One might be tempted to conclude that we have strange weak values and, at the same time, the LGI violated because we may no longer consider the signal photon as a separate system and must consider the signal/meter system together. While this notion helps when interpreting Bell-type experiments, here it might present a difficulty. The weaker the measurement, the less influence the signal has on the meter state.  We should be able to achieve the maximal violation of the LGI in the limit of a separable state but then we would be unable to extract the necessary information to determine the LG correlation function.  The conditions of macroscopic realism and noninvasive measurement are inextricably entwined in the Leggett-Garg inequality and the observation of strange weak values. We were able to test the effect of measurement on the results by controlling the measurement strength but we are unable to separate completely the effect of measurement and the influence, or lack thereof, of realism. A desirable and highly nontrivial extension of the present work would be to design an experiment able to test separately either the macroscopic realism or the noninvasive measurement assumption.

{\bf Acknowledgements} We acknowledge financial support from the Australian Research Council and the IARPA-funded Army Research Office Contract No. W911NF-0397.  We thank Howard Wiseman for helpful discussions.  MG would like to thank the University of Queensland for support and hospitality during his sabbatical visit there.

%

\end{document}